\documentclass[12pt]{article} 
\usepackage{epsfig,amsmath,amssymb,graphics,color,calc,cite}  

\newcommand{\sezione}[2]{  
\refstepcounter{section}\label{#2}  
\setcounter{equation}{0}  
\setcounter{subsection}{0}  
\addcontentsline{toc}{section}  
      {\normalsize\textbf{\thesection.\ #1}}  
\bigskip\bigskip\noindent  
\normalsize\textbf{\thesection.\ #1}\nopagebreak\smallskip\nopagebreak}  
\def\thesection{{\normalsize\arabic{section}}}  
\newcommand{\subsec}[2]{  
\refstepcounter{subsection}\label{#2}  
\addcontentsline{toc}{subsection}  
      {\normalsize\normalfont\textit{\thesubsection.\ #1}}  
\medskip\medskip\noindent  
\normalsize\normalfont 
\textit{\thesubsection. \ #1}\nopagebreak\smallskip\nopagebreak}  
\def\thesubsection{{\normalsize 
{\arabic{section}.\arabic{subsection}}}}  

\newtheorem{theorem}{Theorem}[section]	 
\newtheorem{proposition}[theorem]{Proposition} 
\newtheorem{definition}[theorem]{Definition}	 
\newtheorem{lemma}[theorem]{Lemma}

\newcommand{\cc}[1]{{\mathcal{#1}}} 
\newcommand{\bb}[1]{{\mathbb{#1}}} 
 
\newcommand{\esssup}{\mathop{\rm ess\: sup\,}\limits}
\newcommand{\newatop}[2]{\genfrac{}{}{0pt}{}{#1}{#2}} 
\newcommand{\qed}{\hfill $\Box$\smallskip} 
 
\newcommand{\diam}{\mathrm{diam}} 
\newcommand{\disuno}{\mathrm D} 
\newcommand{\diamuno}{\mathrm{Diam}} 
\renewcommand{\complement}{\mathrm{c}} 
 
\newcommand{\id}{{1 \mskip -5mu {\rm I}}} 
\newcommand{\env}{\cc{Q}} 
\newcommand{\tree}{\bb{T}}

\newcommand{\Es}{Y}

\begin{document} 
\begin{titlepage} 
\par\vskip 1cm\vskip 2em 
 
\begin{center} 
 
{\LARGE Perturbative analysis of disordered\\
\vskip 0.3 cm
Ising models close to criticality}
\par 
\vskip 2.5em \lineskip .5em 
{\large 
\begin{tabular}[t]{c} 
$\mbox{Lorenzo Bertini}^{1} \phantom{m} \mbox{Emilio N.M.\ Cirillo}^{2} 
\phantom{m} \mbox{Enzo Olivieri}^{3}$ 
\\ 
\end{tabular} 
\par 
} 
 
\medskip 
{\small 
\begin{tabular}[t]{ll} 
{\bf 1} & {\it 
Dipartimento di Matematica, Universit\`a di Roma La Sapienza}\\ 
&  Piazzale Aldo Moro 2, 00185 Roma, Italy\\ 
&  E--mail: {\tt bertini@mat.uniroma1.it}\\ 
\\ 
{\bf 2} & {\it 
Dipartimento Me.\ Mo.\ Mat., Universit\`a di Roma La Sapienza}\\ 
&  Via A.\ Scarpa 16, 00161 Roma, Italy\\ 
&  E--mail: {\tt cirillo@dmmm.uniroma1.it}\\ 
\\ 
{\bf 3} & {\it 
Dipartimento di Matematica, Universit\`a di Roma Tor Vergata}\\ 
& Via della Ricerca Scientifica, 00133 Roma, Italy\\ 
& E--mail: {\tt olivieri@mat.uniroma2.it}\\ 
\end{tabular} 
} 
 
\bigskip 
 
 
\end{center} 
 
\vskip 1 em 
 
\centerline{\bf Abstract} 
\smallskip 
We consider a two--dimensional Ising model with random 
i.i.d.\ nearest--neighbor ferromagnetic couplings and no external magnetic 
field. We show that, if  
the probability of supercritical couplings is small enough, the system
admits a convergent cluster expansion with probability one.
The associated polymers are defined on a sequence of increasing scales;
in particular the convergence of the above expansion implies the infinite 
differentiability of the free energy but not its analyticity.
The basic tools in the proof are a general theory of graded cluster expansions
and a stochastic domination of the disorder.
 
\vskip 0.8 em 
 
\vfill 
\noindent 
\textbf{MSC2000.} Primary 82B44, 60K35. 
 
\vskip 0.8 em 
\noindent 
\textbf{Keywords and phrases.} Ising models, Disordered systems, 
         Cluster expansion, Griffiths' singularity. 
 
\bigskip\bigskip 
\footnoterule 
\vskip 1.0em 
{\small 
\noindent 
The authors acknowledge the partial support of Cofinanziamento PRIN. 
\vskip 1.0em 
\noindent 
} 
\end{titlepage} 
\vfill\eject 
 
\sezione{Introduction and main result}{s:int} 
\par\noindent 
In \cite{[BCOabs]} we developed a general theory concerning a
\emph{graded} perturbative expansion for a class of lattice spin systems. 
This theory is useful when the system deserves a multi--scale description  
namely, when a recursive analysis is needed on increasing length scales.
The typical example is provided by a disordered system, like a quenched spin 
glass, having a \emph{good} behavior in average but with the possibility of 
arbitrarily large \emph{bad} regions. Here by good behavior we mean the one of 
a weakly coupled random field. In the bad (i.e.\ not good) regions the system 
can be instead strongly correlated. 
If those bad regions are suitably sparse then the good ones become dominant, 
allowing an analysis based on an iterative procedure.
 
Consider Ising systems described by the following formal Hamiltonian 
which includes the inverse temperature
\begin{equation}
\label{spinglass} 
\cc{H}(\sigma)= 
-\sum_{\newatop{x,y\in\bb{Z}^2}{|x-y|=1}}
J_{x,y}\sigma_x\sigma_y-h\sum_{x}\sigma_x 
\end{equation} 
where $\sigma_x\in\{-1,+1\}$, $h\in\bb{R}$, and $J_{x,y}$ are i.i.d.\ random 
variables. 
A well--known example is the Edwards--Anderson model \cite{[EA]} defined 
by choosing $J_{x,y}$ centered Gaussian random variables with variance 
$s^2$. Let $h=0$; if $s^2$ is small enough, 
so that the probability of subcritical 
couplings is close to one, we expect a weak coupling regime. 
However, in the infinite lattice $\bb{Z}^2$ there are, with probability one,
arbitrarily large regions where the 
random couplings take large positive values giving rise, inside 
these regions, to the behavior of a low--temperature ferromagnetic 
Ising system with long--range order. 

A simpler system is the so called \emph{diluted} Ising model,  
defined by choosing $J_{x,y}$ equal to $K>0$ with probability $q$ and
to zero with probability $1-q$. In this case 
it is possible to show that for $q$ sufficiently small and $K$ sufficiently 
large the infinite--volume free energy is infinitely differentiable 
but not analytical in $h$ \cite{[Gr],[Suto]}. This is a sort of 
infinite--order phase transition called a Griffiths singularity. 
A similar behavior is conjectured for a general distribution of the random 
couplings whenever the probability of supercritical values is sufficiently 
small but strictly positive. 
More precisely, in such a situation, we expect an exponential decay of 
correlations with a non--random decay rate but with a 
random unbounded prefactor. This should imply infinite differentiability of the
limiting quenched free energy, but the presence of arbitrarily large regions 
with supercritical 
couplings should cause the breaking of analyticity.

\medskip
A complete analysis of disordered systems in the Griffiths phase 
is given in \cite{[FI]}, where a powerful and widely applicable perturbative 
expansion has been introduced. The typical applications are 
high--temperature spin glasses and random--field Ising models with large 
variance. 
{}From now on we focus however 
on \emph{ferromagnetic} random Ising systems with
\emph{bounded} couplings $J_{x,y}$ and $h=0$. 
Let $K_{\mathrm{c}}$ be the critical coupling for the standard two--dimensional
Ising model and $K_1<K_{\mathrm{c}}$ be such that for 
coupling constants $J_{x,y}\in[0,K_1]$ the standard high temperature 
cluster expansion is convergent, see e.g.\ \S 20.5.(i) in \cite{[GJ]}.
In the context of \cite{[FI]} a bond $\{x,y\}$ is to be considered \emph{bad}
if the corresponding coupling $J_{x,y}$ exceeds the value $K_1$.
The theory developed in \cite{[FI]} is based on a multi--scale classification
of the bad bonds yielding that, with high 
probability, larger and larger bad regions are farther and farther apart.
In particular there exists a constant $q_1\in(0,1)$ such that 
if $\mathrm{Prob}(J_{x,y}>K_1)\le q_1$ then the system admits a 
convergent graded cluster expansion implying the exponential decay of 
correlations as stated above. 

The aim of the present paper is to analyze disordered systems 
that are weakly coupled only on a sufficiently large scale depending on 
the thermodynamic parameters. In particular we consider random ferromagnetic 
Ising models allowing typical values of the coupling constants arbitrarily 
close to the critical value $K_{\mathrm{c}}$. 
In this case we need a graded cluster expansion based on a 
\emph{scale--adapted} approach. To introduce this notion 
let us take for a while the case of the deterministic ferromagnetic Ising
model with coupling $K$ smaller than the critical value 
$K_{\mathrm{c}}$. The standard high temperature expansions, converging
for coupling smaller than $K_1$, involve perturbations around a 
universal reference system consisting of independent spins.  
In \cite{[O],[OP]} another perturbative expansion
has been  introduced, around a non--trivial model--dependent reference system, 
that can be called \emph{scale--adapted}. Its use is necessary if we 
want to treat perturbatively the system at \emph{any} $K<K_{\mathrm{c}}$ 
since the correlation length can become arbitrarily large close to 
criticality.  
The geometrical objects (polymers) involved in  the
scale--adapted expansion live on a suitable length scale $\ell$ whereas in the 
usual high--temperature expansions they live on scale one. The small parameter 
is no more $K$ but rather the ratio between the correlation length 
at the given $K$ and the length scale $\ell$ at which we analyze the system.  
Of course the smaller is $K_{\mathrm{c}}-K$ 
the larger has to be taken the length $\ell$. 

In the context of the random ferromagnetic Ising model with bounded
interaction, letting $q(b)=\mathrm{Prob}(J_{x,y}>b)$, $b\in\bb{R}_+$,
we prove that there exists a real function $q_0$ such that the
following holds. If for some $b\in[0,K_{\mathrm{c}})$ we have
$q(b)<q_0(b)$, then the system admits, with probability one w.r.t.\
the disorder, a convergent graded cluster expansion implying, in
particular, the exponential decay of correlations with a random
prefactor \cite{[BCOalb]}; such a decay can be proven in a simpler way
by using the methods in \cite{[vDKP]} or in \cite{[Berretti]}, however 
to get the expansion (\ref{tm1-t}) a graded cluster expansion is needed.
The results in \cite{[FI]} can thus be
seen as a special case of the above statement.  We emphasize that
since we consider situations arbitrarily close to criticality, the
first step of our procedure, consisting in the integration over the
good region, is a scale--adapted expansion. In other words, the
minimal length scale involved in the perturbative expansion developed
in the present paper, is not one as in \cite{[FI]}, but rather depends
on the thermodynamic parameters and diverges when approaching the
critical point.  The multi--scale analysis of the bad regions, simpler
than the one in \cite{[FI]}, is achieved by exploiting the
peculiarities of the model. In particular, the basic probability
estimates are deduced via a stochastic domination by a Bernoulli
random field.

\medskip
For $x=(x_1,x_2)\in\bb{R}^2$ we let 
$|x|:=|x_1|+|x_2|$.
The spatial structure is modeled by the two--dimensional lattice  
$\cc{L}:=\bb{Z}^2$ endowed with the distance  
$\disuno(x,y):=|x-y|$.  
We let $e_1$ and $e_2$ be the coordinate unit vectors.  
As usual for $\Lambda,\Delta\subset\cc{L}$ we set  
$\disuno(\Lambda,\Delta):=\inf\{\disuno(x,y),\, x\in\Lambda,\,y\in\Delta\}$ 
and 
$\diamuno(\Lambda):=\sup\{\disuno(x,y),\, x,y\in\Lambda\}$. 
The notation $\Lambda\subset\subset\bb{L}$ means that $\Lambda$ is a 
finite subset of $\cc{L}$.
We let $\cc{E}:=\big\{\{x,y\}\subset\cc{L}:\,\disuno(x,y)=1\big\}$ be the 
collection of \emph{bonds} in $\cc{L}$. 
Given a positive integer $m$ we let $\bb{F}_m$ be the collection of all 
the finite subsets of $\cc{L}$ which can be written as disjoint unions 
of squares with sides of length $m$ parallel to the coordinate axes. 

The single--spin state space is $\cc{X}_0:=\{-1,+1\}$
which we consider endowed with the discrete topology, the  
associated Borel $\sigma$--algebra is denoted by  
$\cc{F}_0$.
The configuration space in $\Lambda\subset\cc{L}$ is defined as  
$\cc{X}_\Lambda:=\cc{X}_0^\Lambda$ and considered 
equipped with the product topology and the corresponding Borel 
$\sigma$--algebra $\cc{F}_\Lambda$. We let $\cc{X}_\cc{L}=:\cc{X}$  
and $\cc{F}_{\cc{L}}=:\cc{F}$. 
Given $\Delta\subset\Lambda\subset\cc{L}$ and  
$\sigma:=\{\sigma_x\in\cc{X}_{\{x\}},\,x\in\Lambda\}\in\cc{X}_\Lambda$,  
we denote by $\sigma_\Delta$ the \emph{restriction} of $\sigma$  
to $\Delta$ namely, $\sigma_\Delta:=\{\sigma_x,\,x\in\Delta\}$. 
Let $m$ be a positive integer and  
let $\Lambda_1,\dots,\Lambda_m\subset\cc{L}$ be pairwise disjoint subsets of  
$\cc{L}$; for each   
$\sigma_k\in\cc{X}_{\Lambda_k}$, with $k=1,\dots,m$, we denote by   
$\sigma_1\sigma_2\cdots\sigma_m$ 
the configuration in $\cc{X}_{\Lambda_1\cup\cdots\cup\Lambda_m}$ such that  
$(\sigma_1\sigma_2\cdots\sigma_m)_{\Lambda_k}=\sigma_k$  
for all $k\in\{1,\dots,m\}$.  
 
A function $f:\cc{X}\to\bb{R}$ is called \emph{local} 
iff there exists $\Lambda\subset\subset\cc{L}$ such that 
$f\in\cc{F}_\Lambda$ namely, $f$ is  
$\cc{F}_\Lambda$--measurable for some bounded set $\Lambda$. 
If $f\in\cc{F}_\Lambda$ we shall 
sometimes misuse the notation by writing $f(\sigma_\Lambda)$ for $f(\sigma)$. 
We also introduce $C(\cc{X})$ the space of continuous functions on  
$\cc{X}$ which becomes a Banach space under the norm  
$\|f\|_\infty:=\sup_{\sigma\in\cc{X}}|f(\sigma)|$; note that the  
local functions are dense in $C(\cc{X})$.  

We let $\cc{J}:=\bb{R}^\cc{E}$, which we consider equipped with its  
Borel $\sigma$--algebra $\cc{B}$. 
We denote by $J_e$, $e\in\cc{E}$, the canonical 
coordinates on $\cc{J}$. 
Let $\bb{P}_0$ be a probability measure on $\bb{R}$ with 
compact support in $\bb{R}_+$ namely, there exists a real $M>0$ such that 
$\bb{P}_0([0,+M])=1$. We define on $(\cc{J},\cc{B})$
the product measure $\bb{P}:=\bb{P}_0^\cc{E}$.
 
Given $\Lambda\subset\subset\cc{L}$, the \emph{disordered 
finite--volume Hamiltonian} is the function  
$H_\Lambda:\cc{X}\times\cc{J}\to\bb{R}$ defined as  
\begin{equation} 
\label{1.2} 
H_\Lambda(\sigma,J):= 
 \sum_{\newatop{\{x,y\}\in\cc{E}:} 
               {\{x,y\}\cap\Lambda\neq\emptyset}}
J_{\{x,y\}}\sigma_x\sigma_y
\end{equation} 
Given $J\in\cc{J}$, 
we define the quenched (finite volume) \emph{Gibbs measure} 
in $\Lambda$, with boundary condition $\tau\in\cc{X}$, as 
the following probability measure on $\cc{X}_\Lambda$. Given 
$\sigma\in\cc{X}_\Lambda$ we set 
\begin{equation} 
\label{gibbs} 
\mu^\tau_{\Lambda,J}(\sigma):= 
  \frac{1}{Z_\Lambda(\tau,J)}  
  \exp\big\{+H_\Lambda(\sigma\tau_{\Lambda^\complement},J)\big\} 
\end{equation} 
where 
\begin{equation} 
\label{poert} 
Z_\Lambda(\tau,J):= 
\sum_{\sigma\in\cc{X}_\Lambda}  
\exp\big\{+H_\Lambda(\sigma\tau_{\Lambda^\complement},J)\big\} 
\end{equation} 
Note that, for notational convenience, we changed the signs in the
definition of the Hamiltonian \eqref{1.2} and in the Gibbs measure
\eqref{gibbs}.

\begin{theorem} 
\label{t:sviluppo} 
Let $K_{\mathrm{c}}:=(1/2)\log(1+\sqrt{2})$ be the
critical coupling of the standard two--dimensional Ising model.
There exists a function $q_0:[0,K_{\mathrm{c}})\to(0,1]$ such that the 
following holds. Suppose that for some $b\in[0,K_{\mathrm{c}})$ 
\begin{equation}
\label{ipop}
q\equiv q(b):=1-\bb{P}_0\big([0,b]\big)\le q_0(b)
\end{equation}
then there exists a positive integer $\ell=\ell(b)$ and 
a set $\bar{\cc{J}}\in\cc{B}$, with $\bb{P}(\bar{\cc{J}})=1$,
such that the following statements hold. 
There exist two families 
$\{\Psi_{X,\Lambda},\Phi_{X,\Lambda}:\cc{X}\times\bar{\cc{J}}\to\bb{R}, 
 \,X\subset\subset\cc{L},\,\Lambda\in\bb{F}_\ell\}$, called 
\emph{effective potential},
such that: 
$\Psi_{X,\Lambda},\Phi_{X,\Lambda}
   \in\cc{F}_{X\cap\Lambda^\complement}\times\cc{B}$ and  
for each $\Lambda,\Lambda'\in\bb{F}_\ell$,   
$X\subset\subset\cc{L}$ such that 
$X\cap\Lambda=X\cap\Lambda'$ one has that 
$\Psi_{X,\Lambda}=\Psi_{X,\Lambda'}$ and 
$\Phi_{X,\Lambda}=\Phi_{X,\Lambda'}$. 
Moreover for each $\Lambda\in\bb{F}_\ell$  
\begin{enumerate} 
\item 
\label{p:tm1} 
for each $(\tau,J)\in\cc{X}\times\bar{\cc{J}}$ we have the convergent
expansion  
\begin{equation} 
\label{tm1} 
  \log Z_{\Lambda}(\tau,J)= 
  \sum_{X\cap\Lambda\neq\emptyset} 
     \left[\Psi_{X,\Lambda}(\tau,J)+\Phi_{X,\Lambda}(\tau,J)\right] 
\end{equation} 
\item 
\label{p:tm3} 
for each $x\in\cc{L}$ there exists a function  
$r_x:\bar{\cc{J}}\to\bb{N}$ such  
that for each $J\in\bar{\cc{J}}$ we have 
$\Psi_{X,\Lambda}(\cdot,J)=0$ for $X\subset\subset\cc{L}$ such that  
$\diam(X)>r_x(J)$ and $X\ni x$; 
\item 
\label{p:tm4} 
there exist reals $\alpha>0$ and $C<\infty$ such that for any 
$J\in\bar{\cc{J}}$
\begin{equation} 
\label{tm2} 
\sup_{x\in\cc{L}} 
 \sum_{X\ni x}e^{\alpha\,\diamuno(X)} 
 \sup_{\Lambda\in\bb{F}_\ell} 
  \|\Phi_{X,\Lambda}(\cdot,J)\|_{\infty} 
<C
\end{equation} 
\end{enumerate} 
\end{theorem} 

In the deterministic case, $J_{x,y}=K$ with 
$K\in[0,K_{\mathrm{c}})$, the expansion (\ref{tm1}) holds with $\Psi=0$
\cite{[O]}. In such a case (\ref{tm2}) implies one of the 
Dobrushin--Shlosman equivalent conditions for \emph{complete analyticity}, 
i.e.\ Condition~IVa in \cite{[DS3]}, see also
equation (2.15) in \cite{[BCOalb]}.
On the other hand, in our disordered setting the family $\Psi$ does not vanish
due to the presence of arbitrarily large regions of strong couplings. 
Nevertheless in item~\ref{p:tm3} we state that the \emph{range}
of the effective potential $\Psi$, although unbounded, is finite with
probability one.
We emphasize that in general the statements in items~\ref{p:tm1}--\ref{p:tm4} 
of Theorem~\ref{t:sviluppo} are not sufficient to deduce complete analiticity.
 
In \cite{[BCOran]} we shall prove a similar result in a more general 
context and, by using the combinatorial approach in \cite{[BCOalb]}, 
deduce an exponential decay of correlations from the convergence of the 
graded cluster expansion.

\sezione{Graded cluster expansion}{s:proth} 
\par\noindent
In this section we prove, relying on some probability estimates
on the multi--scale geometry of the disorder which are discussed in
Section~\ref{s:gra}, Theorem~\ref{t:sviluppo}. 
We follow a classical strategy in disordered systems. Let us fix a 
realization $J\in\cc{J}$ of the random couplings. We first perform a cluster 
expansion in the regions where the model satisfies a strong mixing condition 
implying an effective weak interaction on a proper scale.
We are then left with an effective residual interaction between the regions 
with strong couplings. Since large values of coupling constants have
small probability, the regions of strong couplings are well separated 
on the lattice; we can thus use the graded cluster expansion developed 
in \cite{[BCOabs]} to treat the residual interaction. 

\subsec{Good and bad events}{s:goodandbad}
\par\noindent
Given a positive integer $\ell$, we consider the $\ell$--rescaled
lattice $\cc{L}^{(\ell)}:=(\ell\bb{Z})^2$, which is embedded in
$\cc{L}$ namely, points in $\cc{L}^{(\ell)}$ are also points in
$\cc{L}$, and for each $i\in\cc{L}^{(\ell)}$ we set
\begin{equation}
\label{cubi}
Q_\ell(i):=\{x\in\cc{L}:\,i_1\le x_1\le i_1+\ell-1\;\textrm{and}\;
                          i_2\le x_2\le i_2+\ell-1\} 
\end{equation}
For $i\in\cc{L}^{(\ell)}$ and $b\in[0,K_{\mathrm{c}})$, 
we introduce the bad event
\begin{equation}
\label{badev}
E_i\equiv E^{(\ell),b}_i:=
 \bigcup_{\newatop{e\in\cc{E}:}
                  {e\cap Q_\ell(i)\neq\emptyset}}
  \big\{J_e>b\big\}
\end{equation}
Note that $E_i$ occurs iff in the square $Q_\ell(i)$ there exists a coupling, 
taking into account also the boundary bonds, larger than $b$.
We then define the binary random variable 
$\omega_i\equiv\omega^{(\ell),b}_i:\cc{J}\to\{0,1\}$ as 
\begin{equation}
\label{omega}
\omega_i\equiv\omega^{(\ell),b}_i:=\id_{E^{(\ell),b}_i}
\end{equation}
Given $J\in\cc{J}$ we say that a site $i\in\cc{L}^{(\ell)}$ is 
\emph{good} (resp.\ \emph{bad}) if and only if $\omega_i(J)=0$ 
(resp.\ $\omega_i(J)=1$) and we set 
$\cc{L}_b^{(\ell)}(J):=\{i\in\cc{L}^{(\ell)}:\,\omega_i(J)=0\}$. 

\subsec{On goodness}{s:sogood}
\par\noindent
In this subsection we clarify to which extent the good sites in 
$\cc{L}^{(\ell)}$ are good.
We shall show that for $\ell$ large enough, given
$b\in[0,K_\mathrm{c})$ and $J\in\cc{J}$, on the good part of the
lattice $\cc{L}_b^{(\ell)}(J)$ the quenched disordered model satisfies
a strong mixing condition allowing a nice cluster expansion.

A few more definitions are needed; 
let $i\in\cc{L}^{(\ell)}$ and $k\in\{1,2\}$,
we denote by $P^{i,k}$ the family of all non--empty subsets 
$I\subset\cc{L}^{(\ell)}$ such that for each 
$j\in I$ we have 
$j_k=i_k$ and $j_h\in\{i_h-\ell,i_h,i_h+\ell\}$ for $h=1,2$ and $h\neq k$. 
We set 
$$
I_\pm:=\partial^{(\ell)} I\cap\{j\in\cc{L}^{(\ell)}:\, j_k=i_k\pm\ell\}
$$
where for any $I\subset\cc{L}^{(\ell)}$ we have set 
$\partial^{(\ell)}I:=\{j\in\cc{L}^{(\ell)}\setminus I:\, \disuno(j,I)=\ell\}$.
For $\sigma\in\cc{X}$ we set 
$\sigma_\pm:=\sigma_{\cup_{i\in I_\pm}Q_\ell(i)}$ and 
$\sigma_0:=
\sigma_{(\cup_{i\in I_+\cup I_-}Q_\ell(i))^\complement}$. 
Moreover for each $I\subset\cc{L}^{(\ell)}$ we set 
$\cc{O}_\ell I:=\bigcup_{i\in I} Q_\ell(i)$
and for each $X\subset\cc{L}$ we set 
$\cc{O}^\ell X:=\{i\in\cc{L}^{(\ell)}:\,X\cap Q_\ell(i)\neq\emptyset\}$.

\begin{lemma}
\label{t:bonta}
Let $b\in[0,K_\mathrm{c})$,
there exists an integer $\ell_0=\ell_0(b)$ 
and a real $m_0=m_0(b)>0$ 
such that for each $\ell$ multiple of $\ell_0$, $J\in\cc{J}$, and 
$i\in\cc{L}^{(\ell)}$ we have
\begin{equation}
\label{CC}
\sup_{k=1,2}~
\sup_{I\in P^{i,k}}~
\sup_{\sigma,\zeta,\tau\in\cc{X}}~ 
\left| \frac{Z_{\cc{O}_\ell(I\cap\cc{L}_b^{(\ell)})}
                                         (\sigma_+\sigma_-\tau_0,J) 
             Z_{\cc{O}_\ell(I\cap\cc{L}_b^{(\ell)})}
	                                 (\zeta_+\zeta_-\tau_0,J)} 
            {Z_{\cc{O}_\ell(I\cap\cc{L}_b^{(\ell)})}
	                                 (\sigma_+\zeta_-\tau_0,J) 
             Z_{\cc{O}_\ell(I\cap\cc{L}_b^{(\ell)})}
	                                 (\zeta_+\sigma_-\tau_0,J)} 
-1\right|
<
e^{-m_0\ell}
\end{equation}
\end{lemma}

\smallskip
\par\noindent
{\it Proof.\/} 
The proof, which is based on classical results on (non--disordered)
two--dimensional Ising models adapted to the present 
non--translation--invariant interaction, is organized in three steps. 
Given $b\in [0,K_c)$ we let $\cc{J}_b := [0,b]^\cc{E}\subset\cc{J}$.
We first prove that for each $J\in\cc{J}_b$ there exists a unique 
infinite--volume Gibbs measure w.r.t.\ the {\em local Gibbs specification} 
$Q_{\Lambda,J}(\cdot|\tau):=\mu_{\Lambda,J}^\tau(\cdot)$, 
see \cite[p.~350]{[DS1]} and equation (\ref{gibbs}) above,
$\Lambda\subset\subset\cc{L}$, $\tau\in\cc{X}$.
Then we show that the corresponding infinite--volume two--point correlations
decay exponentially with the distance. From this we finally derive the 
bound (\ref{CC}). 

We consider $\cc X$ endowed with the natural partial ordering
$\sigma\le\sigma'$ iff for any $x\in\cc L$ we have
$\sigma_x\le\sigma'_x$.  Given two probabilities $\mu$, $\nu$ on $\cc
X$ we write $\mu\le \nu$ iff for any continuous increasing (w.r.t.\
the previous partial ordering) function $f$ we have $\mu(f) \le \nu
(f)$. Here $\mu(f)$ denotes the expectation of $f$ w.r.t.\ the measure
$\mu$.

\smallskip
\noindent \emph{Step 1}. 
Let us denote by $+$ (resp.\ $-$) the configuration with all the 
spins equal to $+1$ (resp.\ $-1$). By monotonicity, 
which is a consequence of the FKG inequalities, see 
e.g.\ Theorem~4.4.1 in \cite{[GJ]},
we get that for each $J\in\cc{J}$ and $A\subset\cc{F}$
\begin{equation}
\label{hig1}
\exists\lim_{\Lambda\uparrow\cc{L}}\mu_{\Lambda,J}^\pm(A)=:\mu^\pm_J(A)
\end{equation}
where the limit is taken along an increasing sequence invading
$\cc{L}$. Moreover, again by the FKG inequalities, we
have that any infinite--volume Gibbs measure $\mu_J$ satisfies the
inequalities
\begin{equation}
\label{hig2}
\mu_J^-\le\mu_J\le\mu_J^+
\end{equation}
We now notice that for each $J\in\cc{J}_b$ and $x\in\cc{L}$
\begin{equation}
\label{hig3}
\lim_{\Lambda\uparrow\cc{L}}\mu_{\Lambda,J}^\pm(\sigma_x)=0
\end{equation}
indeed if we let $B\in\cc{J}$ be such that $B_e=b$, $e\in\cc{E}$, 
we have that for each $J\in\cc{J}_b$, $x\in\cc{L}$ and 
$\Lambda\subset\subset\cc{L}$
\begin{equation}
\label{hig4}
\mu_{\Lambda,B}^-(\sigma_x)\le\mu_{\Lambda,J}^-(\sigma_x)
\le0\le
\mu_{\Lambda,J}^+(\sigma_x)\le\mu_{\Lambda,B}^+(\sigma_x)
\end{equation}
where we used the Griffiths inequalities, see
e.g.\ Theorem~4.1.3 in \cite{[GJ]}. By using \cite{[Ons], [LML], [Ru]} and
the FKG inequalities we have that (\ref{hig4}) implies (\ref{hig3})
since $0\le b<K_\textrm{c}$.

Again by FKG, equations (\ref{hig1}) and (\ref{hig3}) imply that for
each $J\in\cc{J}_b$ the infinite--volume Gibbs measure w.r.t.\ the local
specification $Q_{\Lambda,J}(\cdot|\tau)$,
$\Lambda\subset\subset\cc{L}$,
$\tau\in \cc X$ is unique; we denote this measure by $\mu_J$.

\smallskip
\noindent \emph{Step 2}. 
Let $x,y\in\cc{L}$, by the Griffiths' inequalities we have that 
for each $J\in\cc{J}_b$ 
\begin{equation}
\label{hig5}
0\le\mu_J(\sigma_x;\sigma_y)=\mu_J(\sigma_x\sigma_y)\le
\mu_{B}(\sigma_x\sigma_y)
\end{equation}
where we recall $B$ has been defined above (\ref{hig4}).
By using (\ref{hig5}) and classical exact results on the two--dimensional Ising 
model, see e.g.\ \cite{[Bax]}, we have that there
exists a  positive real $C_3(b)<\infty$ such that for any $x,y\in\cc{L}$
\begin{equation}
\label{hig6}
\mu_J(\sigma_x;\sigma_y)\le
\mu_{B}(\sigma_x\sigma_y)\le C_3(b)\,e^{-\disuno(x,y)/C_3(b)}
\end{equation}

\smallskip
\noindent \emph{Step 3}. 
We observe that the argument of the proof of Theorem~2.1 
in \cite{[H]} applies to the present not translationally invariant 
setting. Indeed, it depends only on the Lebowitz inequalities, which
hold true, and the bound \eqref{hig6}.
We thus obtain that there exists a positive real $C_2(b)<\infty$ such
that for each $\tau,\tau'\in\cc{X}$, $J\in\cc{J}_b$,
$\Lambda\subset\subset\cc{L}$, $\Delta\subset\Lambda$, and
$A\in\cc{F}_\Delta$
\begin{equation}
\label{hig7}
|\mu_{\Lambda,J}^\tau(A)-\mu_{\Lambda,J}^{\tau'}(A)|
\le C_2(b)\,e^{-\disuno(\Lambda^\complement,\Delta)/C_2(b)}
\end{equation}
which is, in the terminology introduced in \cite{[MO2]}, the 
\emph{weak mixing} condition for the local specification 
$Q_{\Lambda,J}(\cdot|\tau)$. 
By exploiting the two--dimensionality of the model and 
by using the result in \cite{[MOS]}, we get that there exist an integer 
$\ell_1(b)$ and a positive real $C_1(b)<\infty$ such that for any
$\tau\in\cc{X}$, $J\in\cc{J}_b$, 
$\Lambda\in\bb{F}_{\ell_1(b)}$, $\Delta\subset\Lambda$, 
$x\in\cc{L}\setminus\Lambda$, and 
$A\subset\cc{F}_\Delta$
\begin{equation}
\label{hig8}
|\mu_{\Lambda,J}^\tau(A)-\mu_{\Lambda,J}^{\tau^x}(A)|
\le C_1(b)\,e^{-\disuno(x,\Delta)/C_1(b)}
\end{equation}
where $\tau^x\in\cc{X}$ is given by $\tau^x_x=-\tau_x$ and 
$\tau^x_y=\tau_y$ for all $y\neq
x$, and we recall that the collection of volumes $\bb{F}_\ell$ has been
defined in Section~\ref{s:int}.  
Again in the terminology of \cite{[MO2]}, the bound \eqref{hig8} is
called \emph{strong mixing} condition.
By Corollary~3.2, equations (3.9) and (3.14) in \cite{[O]}, see also 
equation~(2.5.32) in \cite{[OP]},
it implies the statement of the Lemma.  
\qed
 
\subsec{Cluster expansion in the good region}{s:cegood}
\par\noindent
In this subsection we cluster expand the partition function 
over the good part of the lattice. Consider a positive
integer $\ell$ and the $\ell$--rescaled lattice 
$\cc{L}^{(\ell)}=(\ell\bb{Z})^2$. 
We denote by $\disuno_\ell(i,j)=(1/\ell)\disuno(i,j)$ the natural
distance in $\cc{L}^{(\ell)}$ and by $\diamuno_\ell(I):=\sup_{i,j\in
  I}\disuno_\ell(i,j)$ the diameter of a subset
$I\subset\cc{L}^{(\ell)}$. Given $I\subset\cc{L}^{(\ell)}$ and a real
$r>0$, we denote by
$B^{(\ell)}_r(I):=\{j\in\cc{L}^{(\ell)}:\,\disuno_\ell(I,j)\le r\}$
the $r$--neighborhood of $I$.

We associate with each site $i\in\cc{L}^{(\ell)}$ 
the \emph{single--site block--spin configuration space}  
$\cc{X}^{(\ell)}_i:=\cc{X}_{Q_\ell(i)}$.
Given $I\subset\cc{L}$
we consider the \emph{block--spin configuration space}  
$\cc{X}^{(\ell)}_I:=\otimes_{i\in I}\cc{X}_i^{(\ell)}$, 
$I\subset\cc{L}^{(\ell)}$, equipped with the product topology and 
the corresponding Borel $\sigma$--algebra 
$\cc{F}^{(\ell)}_I$.
As before we set $\cc{X}^{(\ell)}:=\cc{X}^{(\ell)}_{\cc{L}^{(\ell)}}$ and 
$\cc{F}^{(\ell)}:=\cc{F}^{(\ell)}_{\cc{L}^{(\ell)}}$.
 
As for the lattices, see the definition just above the Lemma~\ref{t:bonta},
we introduce operators which allow to pack spins and unpack block spins.
We define the \emph{packing} operator $\cc{O}^\ell:\cc{X}\to\cc{X}^{(\ell)}$  
associating with each  
spin configuration $\sigma\in\cc{X}$ the block--spin configuration  
$\cc{O}^\ell\sigma\in\cc{X}^{(\ell)}$ given by  
$(\cc{O}^\ell\sigma)_i:=\{\sigma_x,\,x\in Q_\ell(i)\}$, 
$i\in\cc{L}^{(\ell)}$.  
The \emph{unpacking} operator 
$\cc{O}_\ell:\cc{X}^{(\ell)}\to\cc{X}$ associates  
with each block--spin configuration $\zeta\in\cc{X}^{(\ell)}$ the  
unique spin configuration $\cc{O}_\ell\zeta\in\cc{X}$ such that  
$\zeta_i=\{(\cc{O}_\ell\zeta)_x,\,x\in Q_\ell(i)\}$ for all  
$i\in\cc{L}^{(\ell)}$. 
We remark also that the two operators allow the packing of the  
spin $\sigma$--algebra and the unpacking of the block--spin one namely,  
for each $I\subset\cc{L}^{(\ell)}$ and $\Lambda\subset\cc{L}$ we have  
\begin{equation} 
\label{spisigf} 
\cc{O}_\ell\big(\cc{F}^{(\ell)}_I\big)=\cc{F}_{\cc{O}_\ell I} 
\;\;\;\;\textrm{ and }\;\;\;\; 
\cc{O}^\ell\big(\cc{F}_\Lambda\big)\subset\cc{F}^{(\ell)}_{\cc{O}^\ell\Lambda} 
\end{equation} 
Where in the last relation the equality between the two $\sigma$--algebras 
stands if and only if $\cc{O}_\ell\cc{O}^\ell\Lambda=\Lambda$.  
 
Given $\Delta\subset\subset\cc{L}^{(\ell)}$
we define the \emph{block--spin Hamiltonian}
$H^{(\ell)}_\Delta:\cc{X}^{(\ell)}\times\cc{J}\to\bb{R}$ as  
$H^{(\ell)}_\Delta(\zeta,J):=H_{\cc{O}_\ell\Delta}(\cc{O}_\ell\zeta,J)$
for $\zeta\in\cc{X}^{(\ell)}$ and $J\in\cc{J}$.
The corresponding finite--volume Gibbs measure, with boundary condition 
$\xi\in\cc{X}^{(\ell)}$, is denoted by 
$\mu_{\Delta,J}^{(\ell),\xi}$, the partition function by
$Z_\Delta^{(\ell)}(\xi,J)$ namely,
\begin{equation}
\label{eqpart}
Z_\Delta^{(\ell)}(\xi,J)=Z_{\cc{O}_\ell\Delta}(\cc{O}_\ell\xi,J) 
\end{equation}

Let $J\in\cc{J}$, $\xi\in\cc{X}^{(\ell)}$,
$\Delta\subset\subset\cc{L}^{(\ell)}$, and
recall $\cc{L}_b^{(\ell)}(J)$ has been defined below (\ref{omega}).
In the following Proposition~\ref{t:ipo} we cluster expand 
$Z^{(\ell)}_{\Delta\cap\cc{L}_b^{(\ell)}(J)}(\xi,J)$ and show, in particular,
that Condition~2.1 in \cite{[BCOabs]} is satisfied.
To state the result we need few more definitions. Let 
$\cc{E}^{(\ell)}:=\{\{x,y\}\subset\cc{L}^{(\ell)}:\,\disuno_\ell(x,y)=1\}$ 
the collection of \emph{edges} in $\cc{L}^{(\ell)}$. 
We say that two edges $e,e'\in\cc{E}^{(\ell)}$ 
are connected if and only if $e\cap e'\neq\emptyset$. 
A subset $(V,E)\subset(\cc{L}^{(\ell)},\cc{E}^{(\ell)})$ is said to be 
connected 
iff for each pair $x,y\in V$, with $x\neq y$, there exists in 
$E$ a path of connected edges joining them.  
We agree that if $|V|=1$ then $(V,\emptyset)$ is connected. 
For $X\subset\cc{L}^{(\ell)}$ finite we then set 
\begin{equation}
\label{treedec0} 
\tree_\ell(X):=\inf\big\{|E|,\, (V,E)\subset(\cc{L}^{(\ell)},\cc{E}^{(\ell)}) 
                \textrm{ is connected and } V \supset X\big\} 
\end{equation} 
Note that $\tree_\ell(X)=0$ if $|X|=1$ and for 
$x,y\in\cc{L}^{(\ell)}$ we have $\tree_\ell(\{x,y\})=\disuno_\ell(x,y)$. 
 
\begin{proposition} 
\label{t:ipo} 
Let $b\in[0,K_\mathrm{c})$ and $\ell_0=\ell_0(b)$ as in Lemma~\ref{t:bonta}. 
Then for all integer $\ell$ multiple of $\ell_0$, 
$J\in\cc{J}$, $\xi\in\cc{X}^{(\ell)}$, and 
$\Delta\subset\subset\cc{L}^{(\ell)}$ we have  
\begin{equation} 
\label{ipo} 
\log Z^{(\ell)}_{\Delta\cap\cc{L}_b^{(\ell)}(J)}(\xi,J) 
= 
\sum_{I\cap\Delta\neq\emptyset} 
 V_{I,\Delta}^{(\ell)}(\xi,J) 
\end{equation} 
for a suitable collection of local functions 
$V^{(\ell)}_\Delta:= 
 \{V^{(\ell)}_{I,\Delta}:\cc{X}^{(\ell)}\times\cc{J}\to\bb{R},\,  
                                I\cap\Delta\neq\emptyset\}$ 
satisfying:
\begin{enumerate} 
\item\label{i:cond2.1} 
given $\Delta,\Delta'\subset\subset\cc{L}^{(\ell)}$ if 
$I\cap\Delta=I\cap\Delta'$ 
then $V^{(\ell)}_{I,\Delta}(\cdot,J)=V^{(\ell)}_{I,\Delta'}(\cdot,J)$ 
for any $J\in\cc{J}$; 
\item\label{i:cond2.2} 
$V^{(\ell)}_{I,\Delta}(\cdot,J) 
 \in\cc{F}^{(\ell)}_{I\cap(\Delta\cap\cc{L}_b^{(\ell)}(J))^\complement}$ 
for any $J\in\cc{J}$; 
\item\label{i:cond2.3} 
if $I\cap\big(B^{(\ell)}_6(\Delta)\big)^\complement\neq\emptyset$
then $V^{(\ell)}_{I,\Delta}=0$.
\end{enumerate} 
Moreover, the \emph{effective potential} $V_\Delta^{(\ell)}$ can be bounded 
as follows. There exist reals 
$\alpha_1=\alpha_1(b)>0$, $A_1=A_1(b)<\infty$, and $n_1=n_1(b)<\infty$ such 
that for any $J\in\cc{J}$
\begin{equation} 
\label{ip2} 
\sup_{i\in\cc{L}^{(\ell)}}\, 
 \sum_{I\ni i} e^{\alpha_1\ell\,\tree_\ell(I)} 
  \sup_{\newatop{\Delta\subset\subset\cc{L}^{(\ell)}:} 
                {\Delta\cap I\neq\emptyset}} 
   \|V_{I,\Delta}^{(\ell)}(\cdot,J)\|_{\infty} 
\le A_1\ell^{\,n_1}
\end{equation} 
\end{proposition} 
 
\medskip 
\par\noindent 
{\it Proof of Proposition~\ref{t:ipo}.\/} 
The proof can be achieved by applying the arguments in \cite{[O],[OP]},
where this result is proven with periodic boundary conditions.
We refer to Theorem~5.1 in \cite{[BCObat]} for the modifications needed 
to cover the case of arbitrary boundary conditions and for the stated 
$\ell$--dependence of the bound (\ref{ip2}).
Item~\ref{i:cond2.3} follows from Figures~2 and 3 in \cite{[BCObat]}.
\qed 

\subsec{Geometry of badness}{s:geobad}
\par\noindent
To characterize the sparseness of the bad region we 
follow the ideas developed in \cite{[BCOabs],[BCObat],[FI]}.
 
\begin{definition} 
\label{t:seq-ap} 
We say that two strictly increasing sequences
$\Gamma=\{\Gamma_k\}_{k\ge 1}$ and $\gamma=\{\gamma_k\}_{k\ge 1}$ are
\emph{moderately steep scales} iff they satisfy the following
conditions:
\begin{enumerate} 
\item\label{seq:<}~
$\Gamma_1\ge 2$, and $\Gamma_k<\gamma_k/2$ for any $k\ge 1$;
\item\label{seq:con-dist}~  
for $k\ge 1$ set 
${\displaystyle \vartheta_k:=\sum_{h=1}^k(\Gamma_h+\gamma_h)}$ and 
$\lambda:=\inf_{k\ge 1}(\Gamma_{k+1}/\vartheta_k)$, then $\lambda\ge 5$;
\item\label{seq:covol}~ 
${\displaystyle \sum_{k=1}^\infty\frac{\Gamma_k}{\gamma_k}\le\frac{1}{2}}$ 
\item
\label{seq:con-ano}~ 
${\displaystyle 
a_0:=\sum_{k=0}^{\infty}2^{-k}\log[2(\Gamma_{k+1}+\gamma_{k+1})+1]^2<+\infty}$;
\item
\label{seq:con-icf}~ 
for each $a>0$ we have~ 
${\displaystyle 
\sum_{k=1}^\infty 
\left[2(\vartheta_k+\Gamma_k)+1\right]^2\exp\{-a2^{k-1}\}<\infty
}$.
\end{enumerate} 
\end{definition} 

We remark that items~\ref{seq:con-ano} and \ref{seq:con-icf} 
differ slightly from the corresponding ones in 
Definition~3.1 in \cite{[BCObat]}. This is due to fact that the analysis of 
the geometry of bad sets given in the present paper is based on a 
stochastic domination argument while the one in \S~3 in \cite{[BCObat]} 
depends on mixing properties of the disorder.

\begin{definition} 
\label{t:gentle-ap}  
We say that $\cc{G}:=\{\cc{G}_k\}_{k\ge 0}$, where each  
$\cc{G}_k$ is a collection of  
finite subsets of $\cc{L}^{(\ell)}$, is a 
\emph{graded disintegration of} $\cc{L}^{(\ell)}$ iff 
\begin{enumerate} 
\item\label{i:gd1} 
for each $g\in\bigcup_{k\ge0}\cc{G}_k$  
there exists a unique $k\ge0$, which  
is called the \emph{grade} of $g$, such that $g\in\cc{G}_k$; 
\item\label{i:gd2} 
the collection $\bigcup_{k\ge0}\cc{G}_k$  
of finite subsets of $\cc{L}^{(\ell)}$ is a partition  
of the lattice $\cc{L}^{(\ell)}$ namely, it is a collection of non--empty  
pairwise disjoint finite subsets of $\cc{L}^{(\ell)}$ such that  
\begin{equation} 
\label{part} 
\bigcup_{k\ge 0}\,\bigcup_{g\in\cc{G}_k}g=\cc{L}^{(\ell)}. 
\end{equation} 
\end{enumerate} 
\hfill\break 
Given $\bb{G}_0\subset\cc{L}^{(\ell)}$
and $\Gamma,\gamma$ moderately steep scales, we say that  
a graded disintegration $\cc{G}$ is a \emph{gentle disintegration} of  
$\cc{L}^{(\ell)}$ with respect to $\bb{G}_0,\Gamma,\gamma$ iff 
the following recursive conditions hold: 
\begin{enumerate} 
\setcounter{enumi}{2} 
\item 
\label{gent:null}  
$\cc{G}_0=\big\{\{i\},\,i\in\bb{G}_0\big\}$; 
\item 
\label{gent:diam}  
if $g\in\cc{G}_k$ then $\diamuno_\ell(g)\le\Gamma_k$ for any $k\ge 1$; 
\item 
\label{gent:dist} 
set $\bb{G}_k:=\bigcup_{g\in\cc{G}_k}g\subset\cc{L}^{(\ell)}$,  
$\bb{B}_0:=\cc{L}^{(\ell)}\setminus\bb{G}_0$ and  
$\bb{B}_k:=\bb{B}_{k-1}\setminus\bb{G}_k$, then for  
any $g\in\cc{G}_k$ we have 
$\disuno_\ell(g,\bb{B}_{k-1}\setminus g)>\gamma_k$ for any $k\ge 1$; 
\item 
\label{gent:cas} 
given $g\in\cc{G}_k$, let 
$Y_0(g):=\{j\in\cc{L}^{(\ell)}:\,
             \inf_{i\in\env(g)}[|i_1-j_1|\vee|i_2-j_2|]\le\vartheta_k\}$
where $\env(g)\subset\subset\cc{L}^{(\ell)}$ is the smallest rectangle, 
with axes parallel to the coordinate directions, that contains $g$;
then for each $i\in\cc{L}^{(\ell)}$ we have 
\begin{displaymath}
\varkappa_i:=\sup\big\{k\ge 1:\,\exists g\in\cc{G}_k\textrm{ such that } 
                      Y_0(g)\ni i\}<\infty  
\end{displaymath}
\end{enumerate} 
We call $k$--\emph{gentle}, resp.\ $k$--\emph{bad},
the sites in $\bb{G}_k$, resp.\ $\bb{B}_k$. The
elements of $\cc{G}_k$, with $k\ge 1$, are called $k$--gentle atoms. 
Finally, we set  
$\cc{G}_{\ge k}:= \bigcup_{h\ge k} \cc{G}_h$. 
\end{definition} 

We next state a proposition, whose proof is the topic of 
Section~\ref{s:gra}, which will ensure that for $q$ 
small enough the bad sites of the lattice $\cc{L}^{(\ell)}$ can be classified 
according to the notion of gentle disintegration for suitable scales. 
We note that items~\ref{seq:con-dist} and 
\ref{seq:covol} in Definition~\ref{t:seq-ap}  
force a super--exponential growth of the sequences $\Gamma$ and $\gamma$. 
It is easy to show that, given $\beta\ge8$, the sequences 
\begin{equation} 
\label{sceltaseq-ap} 
\Gamma_k:=e^{(\beta+1)(3/2)^k}\;\;\;\;\;\textrm{ and }\;\;\;\;\; 
\gamma_k:=\frac{1}{8}e^{\beta(3/2)^{k+1}} 
\;\;\;\;\;\textrm{ for } k\ge1 
\end{equation} 
are moderately steep scales in the sense of Definition~\ref{t:seq-ap}.
Given $b\in[0,K_{\mathrm{c}})$, let $\alpha_1(b)$, $A_1(b)$, and 
$n_1(b)$ as in Proposition~\ref{t:ipo}. 
It is easy to show that there exists $\beta_0=\beta_0(b)$ such that  
the scales $\Gamma,\gamma$ in (\ref{sceltaseq-ap}) with $\beta=\beta_0$ 
satisfy the conditions stated in items~1--4 in the hypotheses of 
Theorem~2.5 in \cite{[BCOabs]} for any $\ell$ large enough.
We understand that the 
constants $\alpha$ and $A$ in those items are to be replaced by  
$\alpha_1\ell$ and $A_1\ell^{n_1}$ respectively.

\begin{proposition} 
\label{t:fm}
Given $b\in[0,K_{\mathrm{c}})$ 
let $\Gamma,\gamma$ as in (\ref{sceltaseq-ap}) with $\beta=\beta_0(b)$.
There exist $q_0(b)\in[0,1)$ and 
a multiple $\bar\ell=\bar\ell(b)$ of $\ell_0(b)$, see Lemma~\ref{t:bonta}, 
such that if $q<q_0(b)$, recall (\ref{ipop}), 
then there exists a $\cc{B}$--measurable set $\bar{\cc{J}}\subset\cc{J}$,  
with $\bb{P}(\bar{\cc{J}})=1$, 
such that 
for each $J\in\bar{\cc{J}}$ there exists a gentle disintegration  
$\cc{G}(J)$, see Definition~\ref{t:gentle-ap}, 
of $\cc{L}^{(\bar\ell)}$ with respect to $\cc{L}_b^{(\bar\ell)}(J)$ and 
$\Gamma$, $\gamma$. 
\end{proposition} 

\subsec{Cluster expansion in the bad region}{s:cebad}
\par\noindent
In this subsection we sum over the configurations on  
the bad sites in $\cc{L}^{(\ell)}\setminus\cc{L}_b^{(\ell)}(J)$. 
We show that provided $J$ is chosen in the full $\bb{P}$--measure 
set $\bar{\cc{J}}\subset\cc{J}$, see Proposition~\ref{t:fm},
it is possible to organize the sum 
iteratively using a hierarchy of sparse bad regions of the lattice.  
 
\par\noindent 
{\it Proof of Theorem~\ref{t:sviluppo}.\/} 
We apply Proposition~\ref{t:fm} 
to construct the set $\bar{\cc{J}}$ and, for each $J\in\bar{\cc{J}}$,
the gentle disintegration $\cc{G}(J)$
of $\cc{L}^{(\bar\ell)}$ with respect to $\cc{L}_b^{(\bar\ell)}(J)$ and 
$\Gamma$, $\gamma$ as in (\ref{sceltaseq-ap}) with $\beta=\beta_0(b)$. 
For the seek of simplicity we set $\ell=\bar\ell(b)$ in the sequel of the 
proof. 

Pick $J\in\bar{\cc{J}}$, $\xi\in\cc{X}^{(\ell)}$, and
$\Delta\subset\subset\cc{L}^{(\ell)}$; by applying
Proposition~\ref{t:ipo} we cluster expand $\log
Z^{(\ell)}_{\Delta\cap\cc{L}_b^{(\ell)}(J)}(\xi,J)$.  We get that
Condition~2.1 in \cite{[BCOabs]} holds with effective potential
$V^{(\ell)}_\Delta(\cdot,J)$, $\alpha=\alpha_1\ell$, $A=A_1
\ell^{n_1}$ (here $\alpha_1$, $A_1$, and $n_1$ are the constants
appearing in (\ref{ip2})), and $r=6$.  By applying Theorem~2.5 in
\cite{[BCOabs]} to the lattice $\bb{L}=\cc{L}^{(\ell)}$ and the gentle
disintegration $\cc{G}(J)$ w.r.t.\ $\cc{L}_b^{(\ell)}(J)$, $\Gamma$,
$\gamma$, it follows that there exist functions
$\Psi^{(\ell)}_{I,\Delta}(\cdot,J),\Phi^{(\ell)}_{I,\Delta}(\cdot,J)
\in\cc{F}^{(\ell)}_{I\cap\Delta^\complement}$, with
$I\subset\subset\cc{L}^{(\ell)}$, such that we have the totally
convergent expansion
\begin{equation} 
\label{tm1-t} 
 \log Z^{(\ell)}_{\Delta}(\xi,J)= 
 \sum_{I\cap\Delta\neq\emptyset} 
   \left[\Psi^{(\ell)}_{I,\Delta}(\xi,J)+\Phi^{(\ell)}_{I,\Delta}(\xi,J)\right] 
\end{equation} 
Moreover: $i)$ for each $\Delta,\Delta'\subset\subset\cc{L}^{(\ell)}$ and each  
$I\subset\subset\cc{L}^{(\ell)}$ such that 
$I\cap\Delta=I\cap\Delta'$ we have that 
$\Psi^{(\ell)}_{I,\Delta}=\Psi^{(\ell)}_{I,\Delta'}$ and 
$\Phi^{(\ell)}_{I,\Delta}=\Phi^{(\ell)}_{I,\Delta'}$; 
$ii)$ for $J\in\bar{\cc{J}}$, for $I,\Delta\subset\subset\cc{L}^{(\ell)}$,  
if $\diamuno_\ell(I)>6$   
and there exists no $g\in\cc{G}_{\ge 1}(J)$ such that $\Es_0(g)=X$  
then $\Psi^{(\ell)}_{I,\Delta}(\cdot,J)=0$; $iii)$ 
for each $J\in\bar{\cc{J}}$ we have  
\begin{equation} 
\label{tm2-t} 
\sup_{i\in\cc{L}^{(\ell)}} 
 \sum_{I\ni i}e^{c\alpha_1\ell\diamuno_\ell(I)} 
 \sup_{\Delta\subset\subset\cc{L}^{(\ell)}} 
  \|\Phi^{(\ell)}_{I,\Delta}(\cdot,J)\|_{\infty} 
\le  
1+ 
e^{-c\alpha_1\ell\gamma_1} 
   \Big[  
        \frac{1+e^{-c\alpha_1\ell/4}}{1-e^{-c\alpha_1\ell/4}} 
   \Big]^2
\end{equation} 
where $c=2^{-6}3^{-2}$.
 
To get the expansion (\ref{tm1}) we next pull back 
the $\Psi^{(\ell)}$ and $\Phi^{(\ell)}$ to the original scale.  
We define the family 
$\{\Psi_{X,\Lambda},\Phi_{X,\Lambda}:\cc{X}\times\bar{\cc{J}}\to\bb{R}, 
            \,X\subset\subset\cc{L},\,\Lambda\in\bb{F}_\ell\}$ as follows: 
for each $\tau\in\cc{X}$, $X\subset\subset\cc{L}$, and $\Lambda\in\bb{F}_\ell$ 
we set  
\begin{equation} 
\label{potf0} 
\Psi_{X,\Lambda}(\tau,J):=\Bigg\{ 
\begin{array}{ll} 
{\displaystyle 
 \Psi^{(\ell)}_{I,\cc{O}^\ell\Lambda}(\cc{O}^\ell\tau,J) 
} 
& 
 \textrm{ if } \exists I\subset\cc{L}^{(\ell)}: \cc{O}_\ell I=X 
\\ 
0&\textrm{ otherwise } 
\end{array} 
\end{equation}  
and 
\begin{equation} 
\label{potf1} 
\Phi_{X,\Lambda}(\tau,J):=\Bigg\{ 
\begin{array}{ll} 
{\displaystyle 
 \Phi^{(\ell)}_{I,\cc{O}^\ell\Lambda}(\cc{O}^\ell\tau,J) 
} 
& 
 \textrm{ if } \exists I\subset\cc{L}^{(\ell)}: \cc{O}_\ell I=X 
\\ 
0&\textrm{ otherwise } 
\end{array} 
\end{equation}  
 
Now, the expansion (\ref{tm1}) follows from (\ref{tm1-t}), (\ref{eqpart}), 
(\ref{potf0}), and (\ref{potf1}). The measurability properties of the  
functions $\Psi$ and $\Phi$ follow from (\ref{spisigf}) and  
the analogous properties of the  
functions $\Psi^{(\ell)}$ and $\Phi^{(\ell)}$.  
Item~\ref{p:tm3} follows from item $ii)$ above 
and item~\ref{gent:cas} in Definition~\ref{t:gentle-ap} once we set  
$r_x(J):=\ell[(\Gamma_{\varkappa_x}+2\theta_{\varkappa_x})\vee6]$ 
for each $x\in\cc{L}$ and $J\in\bar{\cc{J}}$, where $\varkappa_x$ is defined
in item~\ref{gent:cas} of Definition~\ref{t:gentle-ap}. 
We finally prove item~\ref{p:tm4}. 
Let $J\in\bar{\cc{J}}$ and set $\alpha:=c\alpha_1$, we have 
\begin{equation} 
\label{stst0} 
\begin{array}{rl} 
{\displaystyle  
 \sup_{x\in\cc{L}} 
 \sum_{X\ni x} e^{\alpha\,\diamuno(X)} 
  \sup_{\Lambda\in\bb{F}_\ell}  
              \|\Phi_{X,\Lambda}(\cdot,J)\|_\infty =  
} 
&  
{\displaystyle  
 \sup_{x\in\cc{L}} 
 \sum_{X\ni x}  
  e^{c\alpha_1\ell\diamuno(X)/\ell} 
  \sup_{\Lambda\in\bb{F}_\ell}  
              \|\Phi_{X,\Lambda}(\cdot,J)\|_\infty 
} \\ 
\vphantom{\Big[} 
= &  
{\displaystyle 
 \sup_{x\in\cc{L}} 
 \sum_{\newatop{I\subset\cc{L}^{(\ell)}:} 
               {\cc{O}_\ell I\ni x}} 
   e^{c\alpha_1\ell\diamuno_\ell(I)} 
  \sup_{\Delta\subset\subset\cc{L}^{(\ell)}}  
    \| 
    \Phi^{(\ell)}_{I,\Delta}(\cdot,J) 
    \|_\infty 
}\\ 
\end{array} 
\end{equation} 
By setting 
\begin{displaymath}
C:=
1+ 
e^{-4\alpha} 
   \Big[  
        \frac{1+e^{-\alpha/4}}{1-e^{-\alpha/4}} 
   \Big]^d 
\end{displaymath}
the bound (\ref{tm2}) follows by using  
(\ref{tm2-t}) and $\gamma_1\ge4$, see item~\ref{seq:<} in  
Definition~\ref{t:seq-ap}. 
\qed

\sezione{Graded geometry}{s:gra} 
\par\noindent
In this section we prove Proposition~\ref{t:fm}.
Recalling the random field $\omega$ has been introduced in (\ref{omega}),
we define $\Omega:=\{0,1\}^{\cc{L}^{(\ell)}}$ and let $\cc{A}$ be the 
corresponding Borel $\sigma$--algebra. We denote by 
$\bb{Q}=\bb{Q}^{(\ell),b}$, a probability 
on $\Omega$, the distribution of the random field $\omega$.
Given $I\subset\cc{L}^{(\ell)}$ we set 
$\cc{A}_I:=\sigma\{\omega_i,\,i\in I\}\subset\cc{A}$. 

Since we assumed the coupling $J_e$, $e\in\cc{E}$,
to be i.i.d.\ random variables, 
the measure $\bb{Q}$ is translationally invariant.
Let us introduce the parameter $p$ which measures the strength 
of the disorder
\begin{equation}
\label{probbadbis}
p:=
   \esssup_{\omega\in\Omega} 
      \bb{Q}\big(\omega_0=1\big|\cc{A}_{\{0\}^\complement}\big) 
\,(\omega)
\end{equation}
where the essential supremum is taken w.r.t.\ $\bb{Q}$. 

We shall first prove that if $p$ is small enough, depending on 
the parameter $a_0$ appearing in item~\ref{seq:con-ano} 
in the Definition~\ref{t:seq-ap}, then we can construct a gentle
disintegration in the sense of Definition~\ref{t:gentle-ap}. 
We finally show that the above condition is met if $q_0(b)$
in (\ref{ipop}) is properly chosen.

Let us first describe an algorithm to construct the family
$\cc{G}$ introduced in Definition~\ref{t:gentle-ap}.  
Given a configuration $\omega\in\Omega$ and $\Gamma$,
$\gamma$ moderately steep scales, we define the following
inductive procedure in a finite volume $\Lambda\subset\subset\cc{L}^{(\ell)}$
which finds the $k$--gentle sites in $\Lambda$.
Set $\bb{G}_0:=\cc{L}_b^{(\ell)}(\omega)$, 
$\cc{G}_0:=\{\{i\},\,i\in\bb{G}_0\}$, and 
$\bb{B}_0:=\cc{L}^{(\ell)}\setminus\bb{G}_0$.
At step $k\ge 1$ do the following:
\texttt{
\begin{enumerate}
\item\label{a1}
$i=1$ and $V=\emptyset$;
\item\label{a2}    
if 
$(\bb{B}_{k-1}\cap\Lambda)\setminus V=\emptyset$ 
then goto \ref{a6};
\item\label{a3} 
pick a point 
$x\in(\bb{B}_{k-1}\cap\Lambda)\setminus V$.   
Set $A=B^{(\ell)}_{\Gamma_k}(x)\cap\bb{B}_{k-1}$   
and $V=V\cup A$;
\item\label{a4} 
if $\diamuno_\ell(A)\le\Gamma_k$ and 
$\disuno_\ell\left(A,\bb{B}_{k-1}\setminus A\right)> \gamma_k$
then $g_k^{i}=A$ and $i=i+1$;
\item\label{a5}
goto \ref{a2};
\item\label{a6}
set $\cc{G}_k :=\{g_k^m,\; m=1,\dots,i-1\}$,  
with the convention $\cc{G}_k=\emptyset$ if $i=1$, 
$\bb{G}_k :=\bigcup_{m=1}^{i-1}g_k^m$, and
$\bb{B}_{k}:=\bb{B}_{k-1}\setminus\bb{G}_k$.  
\end{enumerate} 
}

\par\noindent
Set now $k=k+1$ and repeat the algorithm until 
$\Gamma_k>\diamuno_\ell\Lambda$.

\smallskip 
Let us briefly describe what the above algorithm does. At step $k$ we
have inductively constructed $\bb{B}_{k-1}$, the set of $(k-1)$--bad
sites; we stress that sites in $\cc{L}^{(\ell)}\setminus\Lambda$ may belong to
$\bb{B}_{k-1}$. Among the sites in $\bb{B}_{k-1}\cap\Lambda$ we are
now looking for the $k$--gentle ones.  The set $V$ is used to keep
track of the sites tested against $k$--gentleness. At step \ref{a3} we
pick a new site $x\in\bb{B}_{k-1}\cap\Lambda$ and test it, at step
\ref{a4}, for $k$--gentleness against $\bb{B}_{k-1}$, i.e. including
also bad sites in $\cc{L}^{(\ell)}\setminus\Lambda$.  Note that the families
$\cc{G}_k$ for any $k\ge 1$ are independent on the way in which $x$ is
chosen at step \ref{a3} of the algorithm. Suppose, indeed, to choose
$x\in(\bb{B}_{k-1}\cap\Lambda)\setminus V$ at step \ref{a3} and to
find that $A=B^{(\ell)}_{\Gamma_k}(x)\cap\bb{B}_{k-1}$ is a $k$--gentle
cluster.  Consider $x'\in A$ such that $x'\not= x$ and set
$A':=B^{(\ell)}_{\Gamma_k}(x')\cap\bb{B}_{k-1}$: since $A$ passes the test
against $k$--gentleness at step \ref{a4} of the algorithm, we have
$A\subset A'$. By changing the role of $x$ and $x'$ we get $A=A'$.

After a finite number of operations (bounded by a function of
$|\Lambda|$), the algorithm stops and outputs the family
$\cc{G}_k(\Lambda)$ (note we wrote explicitly the dependence on
$\Lambda$) with the following property. If $g\in\cc{G}_k(\Lambda)$
then $\diamuno_\ell(g)\le\Gamma_k$ and
$\disuno_\ell\left(g,\bb{B}_{k-1}(\Lambda)\setminus g\right) >\gamma_k$.  
Note that $g$ is not necessarily connected.

We finally take an increasing sequence of sets
$\Lambda_i\subset\subset\cc{L}^{(\ell)}$, invading $\cc{L}^{(\ell)}$ and we sequentially
perform the above algorithm.  This means the algorithm for $\Lambda_i$
is performed independently of the outputs previously obtained.
It is easy to show that if
$g\in\cc{G}_k(\Lambda_i)$ then $g\in\cc{G}_k(\Lambda_{i+1})$;
therefore $\cc{G}_k(\Lambda_i)$ is increasing in $i\ge 1$, so that we
can define $\cc{G}_k:=\lim_{i\to\infty}\cc{G}_k(\Lambda_i)
=\bigcup_{i}\cc{G}_k(\Lambda_i)$ and
$\bb{G}_k:=\lim_{i\to\infty}\bb{G}_k(\Lambda_i)=\bigcup_{g\in\cc{G}_k}g$.
Hence,
$\bb{B}_k(\Lambda_i)=\bb{B}_{k-1}(\Lambda_i)\setminus\bb{G}_k(\Lambda_i)=
\cc{L}^{(\ell)}\setminus\cup_{j=0}^{k-1}\bb{G}_j(\Lambda_i)$ is decreasing in
$i\ge 1$, so that $\bb{B}_k:=\lim_{i\to\infty}\bb{B}_k(\Lambda_i)=
\bigcap_i\bb{B}_k(\Lambda_i)$. We also remark that, by construction,
$\{\bb{B}_k,\; k\ge 0\}$ is a decreasing sequence. 
Note that from the construction it follows that it is possible to
decide whether a site $x$ is $k$--gentle by looking only at the
$\omega$'s inside a cube centered at $x$ of radius $\vartheta_k$, as defined
in item \ref{seq:con-dist} of Definition~\ref{t:seq-ap}. 
Hence, see Lemma~3.4 in \cite{[BCObat]}, we have 
the following lemma.

\begin{lemma}
\label{t:mis}
Let $\bb{G}_k$ and $\cc{G}_k$, $k=0,1,\dots$, as constructed
above. Then for each $x\in\cc{L}^{(\ell)}$
\begin{equation}
\label{e:mis} 
\left\{\omega:\, x\in\bb{G}_k(\omega)\right\}
\in\cc{A}_{B^{(\ell)}_{\vartheta_k}(x)}
\end{equation} 
\end{lemma}

\begin{theorem}
\label{t:sexpb}
Let the sequences $\Gamma,\gamma$ satisfy the conditions in items~\ref{seq:<},
\ref{seq:con-dist}, and \ref{seq:con-ano} in 
Definition~\ref{t:seq-ap}. Let also $p<\exp\{-a_0/2\}$ and set 
$a:=-\log p-a_0/2>0 $. Then 
\begin{equation}
\label{sexpb}
\bb{Q}\left(x\in\bb{B}_k\right)\le\exp\{-a\,2^k\} 
\end{equation}
\end{theorem}

\par\noindent
\emph{Remark.\/} 
{}From the previous bound and item~\ref{seq:con-icf}
in Definition~\ref{t:seq-ap},
via a straightforward application of
Borel--Cantelli lemma, see the proof of Theorem~3.3 in 
\cite{[BCObat]} for the details, we deduce the following. 
There exists an $\cc{A}$--measurable set 
$\bar{\Omega}\subset\Omega$,
with $\bb{Q}(\bar\Omega)=1$, such that for each $\omega\in\bar{\Omega}$
there exists a gentle disintegration $\cc{G}(\omega)$, see
Definition~\ref{t:gentle-ap}, of $\cc{L}^{(\ell)}$ with respect to
$\cc{L}_b^{(\ell)}(\omega)$ and $\Gamma$, $\gamma$. 

\smallskip
The first step in proving Theorem \ref{t:sexpb} consists in replacing
the non--product measure $\bb{Q}$ by a Bernoulli product measure with 
parameter $p$. This is a standard argument which we report for completeness.
We consider $\Omega$ endowed with the natural partial ordering
$\omega\le\omega'$ iff for any $x\in\cc{L}^{(\ell)}$ we have
$\omega_x\le\omega'_x$.  
Given two probabilities $Q$, $P$ on $\Omega$ we write $Q\le P$ 
iff for any continuous increasing (w.r.t.\ the previous partial
ordering) function $f$ we have $Q(f)\le P(f)$.

\begin{lemma}
\label{t:stocdom}
Let $\cc{Q}_p$ be the Bernoulli measure on $\Omega$ with marginals
$\cc{Q}_p\left(\omega_x=1\right)=p$ and recall the parameter $p$ has
been defined in \eqref{probbadbis}. 
Then $\bb{Q}\le\cc{Q}_p$.
\end{lemma}

\par\noindent{\it Proof.}\
For $\Lambda\subset\cc{L}^{(\ell)}$ 
we denote by $\bb{Q}_\Lambda$ the marginal of $\bb{Q}$
on $\Omega_\Lambda=\{0,1\}^\Lambda$ and by $Q_p$ the Bernoulli measure on 
$\{0,1\}$, $Q_p(\{1\})=p$. The lemma follows by induction from
\begin{equation}
\label{stocdomin}
\bb{Q}_\Lambda (d\omega_\Lambda) \le 
Q_p (d\omega_x)\bb{Q}_{\Lambda\setminus\{x\}}
  (d\omega_{\Lambda\setminus \{x\}})
\qquad \forall x\in \Lambda,\quad \forall\Lambda\subset\cc{L}^{(\ell)}
\end{equation}
It is easy to show 
$$
\bb{Q}_\Lambda\big(\omega_x=1\big|
  \cc{A}_{\Lambda\setminus\{x\}}\big)= 
\bb{Q}_{\Lambda^\complement}\Big( 
\bb{Q}\big(\omega_x=1\big|\cc{A}_{\{x\}^\complement}\big)\Big) 
\qquad \bb{Q}_\Lambda\textrm{--a.s.}
$$
therefore \eqref{probbadbis} and the translation invariance of $\bb{Q}$ 
imply  
\begin{equation}
\label{pcondiz}
\esssup_{\omega_{\Lambda\setminus\{x\}}}
\bb{Q}_\Lambda\big(\omega_x=1\big|
 \cc{A}_{\Lambda\setminus\{x\}}\big)
\big(\omega_{\Lambda\setminus\{x\}}\big)
\le p
\end{equation}

We next prove \eqref{stocdomin}. Let $f$ be a continuous and increasing 
function on $\Omega_\Lambda$; by taking conditional expectation we have
\begin{equation}
\label{dimstocdomin}
\begin{array}{rcl}
\bb{Q}_\Lambda(f)
&=&
\bb{Q}_{\Lambda\setminus\{x\}}
 \Big(\bb{Q}_\Lambda\big(f\big[\id_{\{\omega_x=1\}}+\id_{\{\omega_x=0\}}
    \big]\big|\cc{A}_{\Lambda\setminus\{x\}}\big)\Big)
\\ 
&=&
{\displaystyle
 \int\! 
 \bb{Q}_{\Lambda\setminus\{x\}}\big(d\omega_{\Lambda\setminus\{x\}}\big)\: 
 \Big\{
 f\big(\omega_{\Lambda\setminus \{x\}} 0_{\{x\}}\big)
}
\\ 
&&
{\displaystyle
\phantom{
 \int\! 
 \bb{Q}_{\Lambda\setminus\{x\}}\big(d\omega
}
 +
 \bb{Q}_\Lambda\big(\omega_x=1
 \big|\cc{F}^{(\omega)}_{\Lambda\setminus\{x\}}\big) 
 \big[f\big(\omega_{\Lambda\setminus \{x\}} 1_{\{x\}}\big)
 - f\big(\omega_{\Lambda\setminus \{x\}} 0_{\{x\}}\big)\big] 
 \Big\}
}
\\
&\le&
{\displaystyle
\int\! 
\bb{Q}_{\Lambda\setminus\{x\}}\big(d\omega_{\Lambda\setminus\{x\}}\big)\: 
\Big\{p\big[f\big(\omega_{\Lambda\setminus \{x\}} 1_{\{x\}}\big) 
        - f\big(\omega_{\Lambda\setminus \{x\}} 0_{\{x\}}\big)\big] 
    + f\big(\omega_{\Lambda\setminus \{x\}} 0_{\{x\}}\big)\Big\}
}
\\ 
&=&
{\displaystyle
\int\!\bb{Q}_{\Lambda\setminus\{x\}} 
 \big(d\omega_{\Lambda\setminus\{x\}}\big) 
Q_p(d\omega_x)f(\omega_\Lambda)
}
\end{array}
\end{equation}
where we used that $f$ is increasing and \eqref{pcondiz} in the
inequality.  
\qed

\begin{lemma}
\label{t:incev} 
For each $x\in\cc{L}^{(\ell)}$ and $k\ge0$ the event 
$\{\omega:\, x\in \bb{B}_k(\omega)\}$ is increasing namely, 
\begin{equation}
\omega\le\omega' 
\Longrightarrow
\bb{B}_k(\omega)\subset\bb{B}_k(\omega')
\label{e:6} 
\end{equation} 
\end{lemma}

\par\noindent{\it Proof.}\
We prove \eqref{e:6} by induction on $k$.
First of all we note that by definition of the natural partial order
on $\Omega$ it holds for $k=0$.  
Let us prove that
\begin{equation}
\bb{G}_{k+1}(\omega')\subset\bigcup_{j=0}^{k+1}\bb{G}_j(\omega)
\label{e:7}
\end{equation}

Let $x\in\bb{G}_{k+1}(\omega')$, then either
$x\in\bigcup_{j=0}^{k}\bb{G}_j(\omega)$ or $x\in\bb{B}_k(\omega)$. In
the former case we are done, in the latter we have that, since
$x\in\bb{G}_{k+1}(\omega')$, there exists a set
$g'\subset\bb{B}_k(\omega')$ such that: $i)$ $x\in g'$; $ii)$
$\diamuno_\ell(g')\le\Gamma_k$; 
$iii)$ $\disuno_\ell(g',\bb{B}_k(\omega')\setminus
g')>\gamma_k$. Set now $g:=g'\cap\bb{B}_k(\omega)$, by the
inductive hypotheses it is easy to verify that $g$
satisfies the three properties above with $\omega'$ replaced by
$\omega$. Hence $x\in\bb{G}_{k+1}(\omega)$. From 
\eqref{e:7} and the induction hypotheses we get 
$\bigcup_{j=0}^{k+1}\bb{G}_j(\omega')\subset\bigcup_{j=0}^{k+1} 
\bb{G}_j(\omega)$.
Since
$\bb{B}_{k+1}(\omega)=\bb{B}_0(\omega)
  \setminus\bigcup_{j=0}^{k+1}\bb{G}_j(\omega)$,
we have proven \eqref{e:6} with $k$ replaced by $k+1$. 
\qed

The key step in proving Theorem \ref{t:sexpb} is the following
recursive estimate on the degree of badness.

\begin{lemma}
\label{t:recatt} 
Let $\Gamma,\gamma$ satisfy the conditions in items~\ref{seq:<},
\ref{seq:con-dist}, and \ref{seq:con-ano} in Definition~\ref{t:seq-ap}
and set $\psi_k := \cc{Q}_p \left( x\in \bb{B}_k \right)$, note
$\psi_k$ is independent of $x$ by translational invariance, and
$A_k(x):=B^{(\ell)}_{\gamma_k+\Gamma_k}(x)\setminus B_{(\Gamma_k-1)/2}(x)$. 
Then
\begin{equation}
\label{recatt} 
\psi_{k+1} \le |A_{k+1}| \psi_k^2
\end{equation}
where $|A_k|=|A_k(x)|$ does not depend on $x$. 
\end{lemma}

\par\noindent{\it Proof.}\
By recalling the definition of the $k$--bad set $\bb{B}_k$ we have 
\begin{equation}
\label{e:1}
\left\{x\in\bb{B}_{k+1}\right\}=
\left\{x\in\bb{B}_k\right\}\cap 
\left\{x\not\in\bb{G}_{k+1}\right\}
\end{equation}  
On the other hand, by the construction of the ($k+1$)--gentle sites,  
\begin{equation}
\label{e:2}
\left\{x\in\bb{B}_k\right\}\cap 
\left\{x\not\in\bb{G}_{k+1}\right\}
\subset
\left\{x\in\bb{B}_k\right\}\cap 
\left\{\exists\,y\in A_{k+1}(x):\; y\in\bb{B}_k\right\}
\end{equation}
indeed, given $\bb{B}_k$, if there were no $k$--bad site in the annulus
$A_{k+1}(x)$ then $x$ would have been $(k+1)$--gentle. {}From
\eqref{e:1} and \eqref{e:2}
\begin{equation}
\label{e:3}
\begin{array}{rl}
{\displaystyle
\psi_{k+1}=\cc{Q}_p\left(x\in\bb{B}_{k+1}\right)\le
}
&
{\displaystyle
\cc{Q}_p
\bigg(
\bigcup_{y\in A_{k+1}(x)}
\left\{x\in\bb{B}_k\right\}
\cap
\left\{y\in\bb{B}_k\right\}
\bigg)
}
\\
&\\
{\displaystyle
\le
}
&
{\displaystyle
\sum_{y\in A_{k+1}(x)}
\cc{Q}_p
\big(
\left\{x\in\bb{B}_k\right\}
\cap
\left\{y\in\bb{B}_k\right\}
\big)
}
\\
&\\
{\displaystyle
=
}
&
{\displaystyle
\sum_{y\in A_{k+1}(x)}
\cc{Q}_p
\big(
\left\{x\in\bb{B}_k\right\}
\big)
\cc{Q}_p
\big(
\left\{y\in\bb{B}_k\right\}
\big)=|A_{k+1}|\psi_k^2 
}
\\
\end{array}
\end{equation} 
where in the last step, we used \eqref{e:mis}, the definition of
$A_k(x)$, item~\ref{seq:con-dist} in Definition~\ref{t:seq-ap}, the
product structure of the measure $\cc{Q}_p$ and its translation
invariance. \qed

\medskip
\par\noindent
{\it Proof of Theorem~\ref{t:sexpb}.\/}
By Lemmata \ref{t:stocdom} and \ref{t:incev} it is enough to prove the
bound \eqref{sexpb} for the Bernoulli measure $\cc{Q}_p$.

Let $f_k:=-\log\psi_k$ and $b_k:=\log|A_{k+1}|$, where $\psi_k$ and 
$A_k$ have been defined in Lemma \ref{t:recatt}. Then by iterating 
\eqref{recatt} and using item \ref{seq:con-ano} in 
Definition~\ref{t:seq-ap}, we get 
\begin{equation} 
f_{k+1}\ge
2f_k-b_k
\ge\cdots\ge
2^{k+1}f_0-2^k\sum_{j=0}^k2^{-j}b_j
\ge
2^{k+1}f_0-2^ka_0
= 2^{k+1}a
\label{e:4} 
\end{equation}
where we recall that $a=-\log p -a_0/2=f_0-a_0/2>0$.
\qed 
 
\medskip
Recall $q$ has been defined in 
(\ref{ipop}). To gently disintegrate the lattice $\cc{L}^{(\ell)}$ by means of 
Theorem~\ref{t:sexpb}, we 
need a bound on the badness parameter $p$, see (\ref{probbadbis}), 
in terms of $q$. 

\begin{lemma}
\label{t:probbad}
Recall $p$ has been defined in (\ref{probbadbis}) and 
$q$ in (\ref{ipop}); then
\begin{equation}
\label{probbad2}
p
\le
1-(1-q)^{2\ell(\ell-1)}+4\frac{1-(1-q)^\ell}{1-(1-q)^{2\ell(\ell+1)}}
\end{equation}
\end{lemma}
\smallskip
\par\noindent
{\it Proof.\/} For each $i\in\cc{L}^{(\ell)}$ we define the five events 
$E_i^0$, $E_i^{1,\pm}$, and $E_i^{2,\pm}$:
\begin{equation}
\label{evint}
E_i^{0}:=
 \bigcup_{\newatop{e\in\cc{E}:}{e\subset Q_\ell(i)}}
  \big\{J_e>b\big\}
\;\;\;\;\textrm{and}\;\;\;\;
E_i^{s,\pm}:=
 \bigcup_{\newatop{e\in\cc{E}:\,e\cap Q_\ell(i)\neq\emptyset,}
                  {e\cap Q_\ell(i\pm\ell e_s)\neq\emptyset}}
  \big\{J_e>b\big\}
\end{equation}
where $s=1,2$ and we 
recall $e_1$ and $e_2$ are the coordinate unit vectors in $\cc{L}$. 
By using the equality 
\begin{displaymath}
E_i=E_i^0\cup E_i^{1,-}\cup E_i^{2,+}\cup E_i^{1,+}\cup E_i^{2,-}
\end{displaymath}
and the product nature of $\bb{P}$, 
we have that 
\begin{equation}
\label{pca1}
\begin{array}{rcl}
\bb{P}(E_i|\{\omega_j=a_j\}_{j\neq i})
&\le&
\bb{P}(E_i^0|\{\omega_j=a_j\}_{j\neq i})
\\
&&
{\displaystyle 
+\sum_{s=1}^2
\bb{P}(E_i^{s,+}|\{\omega_j=a_j\}_{j\neq i})
+\sum_{s=1}^2
\bb{P}(E_i^{s,-}|\{\omega_j=a_j\}_{j\neq i})
}
\\
&=&
\bb{P}(E_i^0)+4\bb{P}(E_i^{1,-}|\omega_{i-\ell e_1}=a_{i-\ell e_1})
\end{array}
\end{equation}
Since $E_i^{1,-}\cap\{\omega_{i-\ell e_1}=0\}=\emptyset$, the l.h.s.\
of (\ref{pca1}) can be bounded uniformly in $\{a_j\}_{j\neq i}$ by 
$\bb{P}(E_i^0)+4\bb{P}(E_i^{1,-})/\bb{P}(\omega_{i-\ell e_1}=1)$. The 
lemma then follows by a straightforward computation.
\qed

\smallskip
\par\noindent
{\it Proof of Proposition~\ref{t:fm}.\/}
Let the functions $q_0:[0,K_{\mathrm{c}})\to(0,1]$ and
$\bar\ell:[0,K_{\mathrm{c}})\to[0,\infty)$ be constructed as follows.
\begin{enumerate}
\item
By Lemma~\ref{t:bonta}, given $b\in[0,K_{\mathrm{c}})$, we find 
$\ell_0(b)$ and $m_0(b)$ such that (\ref{CC}) holds.
\item
By Proposition~\ref{t:ipo} we find 
$\alpha_1(b)$, $A_1(b)$, and $n_1(b)$ such that the bound (\ref{ip2}) 
holds. 
\item
Choose $\beta_0(b)$ as below (\ref{sceltaseq-ap}) and let the 
scales $\Gamma,\gamma$ be as in (\ref{sceltaseq-ap}) with $\beta=\beta_0(b)$.
\item
Compute $a_0(b)$ in item~\ref{seq:con-ano} in Definition~\ref{t:seq-ap}.
\item
\label{i:cinque}
Let $q_0(b)$ and $\bar\ell(b)$, with $\bar\ell(b)$ multiple of
$\ell_0(b)$ such that
\begin{displaymath}
1-(1-q)^{2\bar\ell(b)(\bar\ell(b)-1)}
+4\frac{1-(1-q)^{\bar\ell(b)}}{1-(1-q)^{2\bar\ell(b)(\bar\ell(b)+1)}}
<\exp\{-a_0(b)/2\}
\end{displaymath}
for any $q\le q_0(b)$. 
\end{enumerate}
Step~\ref{i:cinque} is possible because for each $\bar\ell(b)$ fixed the 
l.h.s.\ of the inequality above converges to $2/(\ell+1)$ as $q\to0$. 

By applying Lemma~\ref{t:probbad}, 
Theorem~\ref{t:sexpb}, and the remark following it
we conclude the proof of the proposition.
\qed


 
 
 
\end{document}